\documentclass[12pt]{iopart}
\usepackage{graphicx}

\usepackage{amssymb}
\begin{document}

\title[Dynamics of the Pionium with the Density Matrix Formalism]%
{Dynamics of the Pionium with the Density Matrix Formalism}

\author{L Afanasyev$^\S$, C Santamarina$^\dagger$
\footnote[7]{now at CERN, CH-1211 Gen\`eve 23 (Switzerland)}, 
A Tarasov$^\S$ and O Voskresenskaya$^\S$}
\address{$\dagger$ Institut f\"ur Physik, Universit\"at Basel,
  4056 Basel, Switzerland}
\address{$\S$ Joint Institute for Nuclear Research, 141980 Dubna,
Moscow Region, Russia}

\ead{Cibran.Santamarina.Rios@cern.ch.}

\begin{abstract}
The evolution of pionium, the $\pi^+ \pi^-$ hydrogen-like atom while
passing through mater is
solved within the density matrix formalism in the first Born approximation.
We compare the influence on the pionium lifetime between the standard
break-up probability calculations and the more precise picture of the
density matrix formalism accounting for interference effects.
We focus our general result in the particular conditions of the DIRAC
experiment at CERN.
\end{abstract}

\pacs{34.50.-s,32.80.Cy,36.10.-k,13.40.-f}

\submitto{\JPB}

\maketitle

\section{Introduction}

The evolution of pionium, the hydrogen-like atom formed by a $\pi^+ \pi^-$ pairs,
in a material target has been thoroughly studied in the recent
years~\cite{mrow,afan,hala,heim,schu,sant2,sant} due to its
crucial implications in the DIRAC-PS 212 experiment~\cite{prop}. This experiment is devoted
to measure the lifetime of pionium, intimately linked to the strong interaction scattering
lengths as we will see in~\sref{eveq}, testing the predictions of the Chiral Perturbation
Theory on these magnitudes.

The transport of pionium in mater has been always treated using a classic
probabilistic picture neglecting the quantum mechanics interference between
degenerated states with the same energy. In the case of hydrogen-like atoms
this is of particular importance since the accidental degeneracy of the
hamiltonian increases the amount of states among which the interference can be significative.

In~\cite{voskre} the density matrix formalism has been used to propose a new set of equations
for the pionium evolution accounting for the interference effects. In this
work we have solved these equations and analyzed the consequences for the framework
of DIRAC experiment.

\section{The Problem of Pionium in Matter}
\label{eveq}

Due to the short lifetime
of the pion, pionium, the hydrogen-like $\pi^+ \pi^-$ atom
can not be produced at rest in the laboratory frame. 
However, pionium can be originated in 
collisions of high-energy projectiles with a fixed target. 
The production cross section is given in~\cite{neme}:
\begin{equation}
\frac{\rmd\sigma^A_{i}}{\rmd\vec{P}} =
(2\pi)^3 \left|\psi_{i}(0)
\right|^2\vphantom{\Bigg|_{\vec p}}
\frac{E}{M}
\frac{\rmd\sigma^0_{s}}{\rmd\vec{p}\,\rmd\vec{q}}
\Bigg|_{\vec{p}=\vec{q}=\vec{P}/2},
\end{equation}
where the rightmost term accounts for
the production of $\pi^+$ and $\pi^-$ pairs at equal momenta ($\vec{p}=\vec{q}$).

The state of pionium is defined by
the center of mass momentum $\vec{P}$ and the eigenstate quantum numbers,
$n_i$, $l_i$ and $m_i$, of the hydrogen-like hamiltonian. For simplicity, in this work we have
chosen to work with monochromatic atoms of 4.6~GeV/c, the mean
value of laboratory momentum of pionium in DIRAC, moving in the
$z$ axis direction. The effect of using the experimental
pionium laboratory momentum spectrum is small as shown in~\cite{sant}.
The yield of a particular
state is proportional to its wave function squared at the origin.
It has been shown~\cite{kura} that the effect of the strong
interaction between the two pions of the atom significantly modifies
$|\psi_{i}(0)|$ in comparison to the pure Coulomb wave function.
However, the ratio between the production rate in different states
has been demonstrated to be kept as for the Coulomb wave functions~\cite{amir}.
Thus, considering that the Coulomb functions obey
\begin{equation}
\label{wfat0}
\left|\psi^{(\mathrm{C})}_{i}(0)\right|^2 = \left\{
\begin{array}{cl}
0 & \textrm{if $l_i \neq 0$,} \\
\displaystyle\frac{(\alpha M_{\pi}/2)^3}{\pi n_i^3} & \textrm{if $l_i=0$,}
\end{array}
\right.
\end{equation}
we see that only S states are created following the $1/n_i^3$ law.

The atom moves in
a fixed thickness target disposed in the $Oz$ axis and considered infinite
in the transverse $(x,y)$ coordinates. The target is made of a chemically pure
material like Nickel, Platinum or Titanium. Our goal is to know 
the population probability of every bound state as a function
of the position in the target, $z$, and from this extract other results as the break-up probability. Usually
a classical approach
is used to solve this problem~\cite{afan,sant2}. It consists of considering the total $\sigma_{i}^{tot}$
and transition between two discrete states $\sigma_{i,l}$ cross sections
for a pionium-target atom scattering and apply the probabilistic evolution equation:
\begin{equation}
\label{eq:cl}
\frac{\rmd P_i(z)}{\rmd z} = 
- \frac{1}{\beta \gamma} \Gamma_i P_i(z) - n_0 \sum_{l} c_{i,l} P_l(z),
\end{equation}
where $P_i(z)$ is the classical probability for the atom to be in the $i$ state, $\beta \gamma = 16.48$
the Lorentz center of mass
to laboratory factor for $P=4.6$~GeV/c, $n_0$ is the number of target atoms per unit of volume,
and $c_{i,l}$ are the transition coefficients.

The value of $n_0$ is a function of the density of the target, $\rho$, the Avogadro number, $N_0$,
and the atomic mass of the target atoms, $A$:
\begin{equation}
n_0 = \frac{\rho N_0}{A},
\end{equation}
while the transition coefficients depend on the pionium-target atom
cross sections as:
\begin{equation}
\label{eq:cil}
c_{i,l} = \delta_{i,l} \sigma_{i}^{tot} - \sigma_{i,l}.
\end{equation}

The pionium decay is strongly dominated (BR $> 99\%$~\cite{neme}) by the $\pi^+ \pi^- \rightarrow \pi^0 \pi^0$
reaction.
Taking this into account, the width of the $i$ state is proportional to the isospin
0 and isospin 2 pion-pion scattering lengths difference~\cite{gasse}:
\begin{equation}
\label{tau}
\Gamma_i = \frac{16\pi}{9}\frac{\sqrt{M_{\pi}^2-M_{\pi^0}^2
-\frac{1}{4} M_{\pi}^2\,\alpha^2}}{M_{\pi}} (a_0^0 -a_0^2)^2
(1+\delta_{\Gamma}) \left|\psi^{(\mathrm{C})}_{i}(0)\right|^2 \; ,
\end{equation}
where $M_{\pi}$ and $M_{\pi^0}$ are the masses of the charged and the neutral
pion and $\delta_{\Gamma}=0.058$ the Next to Leading Order correction that
includes the effect of the strong interaction between the two pions.
Of course, the width of a state holds $\Gamma_i=\tau_i^{-1}$ where
$\tau_i$ is the corresponding lifetime of the state. Due to~\eref{wfat0}
we can see that pionium only decays from S states and the
lifetime of any S state is related to the lifetime of the ground state:
\begin{equation}
\tau_{n00} = n^3 \tau.
\end{equation}
The lifetime of pionium is hence the only parameter to be inputed in the evolution
equation and can be related to any of its outputs. In particular we will link it
to the break-up probability. The experimental result of DIRAC will be used to
test with $5\%$ accuracy the accurate 
Chiral Perturbation Theory prediction of $a_0^0-a_0^2=0.265\pm 0.004$
which leads to the lifetime value of
$\tau = (2.9\pm 0.1)\cdot 10^{-15}$~s~\cite{cola}.

\section{The Density Matrix Evolution Equation}

Equation~\eref{eq:cl} has been accurately solved obtaining the eigenvalues
and the eigenvectors~\cite{afan} and also with Monte Carlo~\cite{sant} for the
bound states with $n<8$, which is enough to precisely calculate
the break-up probability as explained in~\cite{sant}.
However, the work of Voskresenskaya~\cite{voskre} demonstrates that
the use of the classic probabilistic picture might be inaccurate.
This is because~\eref{eq:cl} neglects the quantum interference between the pionium states
during their passage through the target.

A more precise description of the system dynamics is given in terms of the
density matrix $\rho_{ik}$. The evolution equation in this formalism is given by~\cite{voskre}:
\begin{equation}
\label{eq:dedm}
\frac{\partial \rho_{ik}}{\partial z} = 
\frac{1}{\beta \gamma} \left[ i(\varepsilon_k -\varepsilon_i) - \frac{1}{2}
(\Gamma_i + \Gamma_k)\right] \rho_{ik}(z) - 
n_0 \sum_{l,m}  \Omega_{ik,lm} \rho_{lm}(z)
\end{equation}
where $\varepsilon_k$ indicates the bounding energy of the $k$ state and
$\Omega_{ik,lm}$ stands for the transition coefficients matrix.
This equation reduces to~\eref{eq:cl}, identifying $\rho_{ii}(z)=P_i(z)$, if
the $\Omega_{ik,lm}$ crossed
terms obeying $i \ne k$ or $l \neq m$ were zero.

The goal of this work is to solve this equation and determine how it
corrects~\eref{eq:cl} for the particular conditions of DIRAC experiment, namely
for the result of the break-up probability.

\section{The Matrix Elements}

To calculate the matrix
elements $c_{i,l}$ and $\Omega_{ik,lm}$ we have applied the coherent pure electrostatic
first Born approximation approach. Even though it is
known that relativistic and multiphoton exchange must be accounted to achieve 
the precision of $1\%$~\cite{sant} our goal was to check wether quantum interference
is a relevant effect. For this we will show that pure electrostatic first Born approximation is enough.

The expression for the pionium-target cross sections in the electrostatic first Born
approximation, used in the
classical picture, was obtained by S. Mr\'owczy\'nski time ago~\cite{mrow}:
\begin{equation}
\label{eq:sti}
\sigma_{i}^{tot}  =  \frac{2}{\beta^{2}} \int
\left| U(q) \right|^{2} \left[1-F_{i}^{i}(\vec{q})
\right] \rmd^2 q,
\end{equation}
\begin{equation}
\label{eq:sil}
\sigma_{i,l} =  \frac{1}{\beta^{2}}
\int \left| U(q) \right|^{2} \left|
F_{i}^{l}\left(\frac{\vec{q}}{2}\right) -
F_{i}^{l}\left(-\frac{\vec{q}}{2}\right) \right|^{2} 
\rmd^2 q,
\end{equation}
where $q$ is the transferred momentum between the target and the pionic atoms.
The cross section does only depend on the two transverse coordinates of the momentum 
due to the symmetry of the collision with respect to the scattering axis.
We have chosen the Fourier transform of the target atom potential $U(q)$ 
to be the Moli\`ere parameterization for the solution
of the Thomas-Fermi equation~\cite{moliere}:
\begin{equation}
\label{eq:pot}
U(q) = 4\pi Z \alpha 
\left(\frac{0.35}{q^2+q_0^2}+\frac{0.55}{q^2+16 q_0^2}
+\frac{0.10}{q^2+400 q_0^2}\right) \quad q_0 = \frac{0.3 Z^{1/3}}{0.885 a_0},
\end{equation}
being $a_0=0.529 \times 10^{-28}$ cm the Bohr radius of Hydrogen, $\alpha$
the fine structure constant and $Z$ the atomic number of the target atoms.
The $F_{i}^{l}(q)$ are the pionium form factors:
\begin{equation}
\label{eq:ff}
F_{i}^{l}(\vec{q}) = \int \psi_l^*(\vec{r}) e^{i \vec{q}\vec{r}} \psi_i^*(\vec{r}) \rmd \vec{r},
\end{equation}
calculated in~\cite{afan} and~\cite{hala}.
In this work we shall use the code of~\cite{sant3} based on the result of~\cite{afan}.

The equivalent of~\eref{eq:cil} for the $\Omega_{ik,lm}$ elements in
the density matrix formalism is given by:
\begin{equation}
\Omega_{ik,lm} = \Omega^{(1)}_{ik,lm} - \Omega^{(2)}_{ik,lm},
\end{equation}
where:
\begin{eqnarray}
\Omega^{(1)}_{ik,lm} = 
& \frac{\delta_{k,m}}{2\beta^2}  \int
\left| U(q) \right|^{2} \left[2 \delta_{i,l}-F_{i}^{l}(\vec{q})-F_{i}^{l}(-\vec{q})
\right] \rmd^2 q +  \nonumber \\ 
& + \frac{\delta_{i,l}}{2\beta^2} \int
\left| U(q) \right|^{2} \left[2 \delta_{k,m}-F_{k}^{m}(\vec{q})-F_{k}^{m}(-\vec{q})
\right] \rmd^2 q,
\label{eq:om1}
\end{eqnarray}
plays the role of the total cross section, while
\begin{eqnarray}
\Omega^{(2)}_{ik,lm} = \frac{1}{\beta^2} 
  \int
 \left| U(q) \right|^{2} & \left[
F_{i}^{l}\left(\frac{\vec{q}}{2}\right)-
F_{i}^{l}\left(-\frac{\vec{q}}{2}\right) \right] \times \nonumber \\
&\times \left[F_{k}^{m}\left(\frac{\vec{q}}{2}\right)-
F_{k}^{m}\left(-\frac{\vec{q}}{2}\right) \right]^{*} \rmd^2 q,
\label{eq:om2}
\end{eqnarray}
would be the analogue of the transition cross section. In fact
$\Omega^{(1)}_{ik,lm}$ becomes the total cross section if
$i=k=l=m$ and $\Omega^{(2)}_{ik,lm}$ 
the transition cross section if $i=k$ and $l=m$.

Equations~\eref{eq:om1} and~\eref{eq:om2} are our main tool for the numerical calculations
and their development from the original
formulas of~\cite{voskre} can be followed in~\ref{app:omega}.

\subsection{Selection Rules and Transition Elements Examples}

As pointed out in~\cite{voskre}, and due to the properties of the form
factors under the parity transformation, the $\Omega_{ik,lm}$ coefficients are different
from zero only if:
\begin{equation} 
\label{eq:quan}
m_i-m_k-m_l+m_m=0\,,\quad l_i-l_k-l_l+l_m=2s
\end{equation}
where we should remember that $m_{i(k,l,m)}$ and $l_{i(k,l,m)}$
are the magnetic and orbital quantum numbers of the states
$|i(k,l,m)\rangle$. The index $s$ is an arbitrary integral number.

For the election of the $Oz$ axis as the quantization axis
the transitions between states
of different $z$-parity are strongly suppressed~\cite{afan}. This means that
only states with even $l-m$ will be populated since pionium is
produced in S states only. This, together with~\eref{eq:quan}
means that:
\begin{equation}
\label{eq:rel0}
\rho_{ik}(z)\ne 0\quad\mbox{if}\quad m_i=m_k\,,
\quad l_i=l_k+2s.
\end{equation}
This rule could be broken by the complex coefficient in~\eref{eq:dedm}:
\[
i\frac{(\varepsilon_k - \varepsilon_i)}{\beta \gamma}, 
\]
which produces an oscillatory term in the solutions.
However, for the ground and lowest excited states the condition:
\[
n_0 \left|\Omega_{ik,ik}\right| \ll \frac{|\varepsilon_k - \varepsilon_i|}{\beta \gamma} 
\]
holds and the $\rho_{ik}(z)$ solution oscillates many times in a small interval, compared to the
electromagnetic transition range (given by $n_0 \left|\Omega_{ik,ik}\right|$) and can be considered
to average as zero:
\[ \rho_{ik}(z) \approx 0. \]
There is an exception if the $i$ and $k$ states belong to the same shell since the energy
of the hydrogen-like system does only depend on the principal quantum number. In this case
$\varepsilon_k - \varepsilon_i =0$. Hence, for the low energy states, we can complete relation~\eref{eq:rel0} as:
\begin{equation}
\rho_{ik}(z)\ne 0\quad\mbox{if}\quad \varepsilon_i=\varepsilon_i\, (n_i=n_k) \,, \quad m_i=m_k\,, \quad l_i=l_k+2s.
\end{equation}

However, if the principal quantum numbers of the $i$ and $k$ states hold $n_{i,k} \gtrsim 6$, then
\[
n_0 \left|\Omega_{ik,ik}\right| \sim \frac{|\varepsilon_k - \varepsilon_i|}{\beta \gamma}
\]
and the solution for $\rho_{ik}(z)$ is not zero even though $i$ and $k$ are not states from the same shell.

\begin{figure}
\centerline{\includegraphics[width=0.45\textwidth]{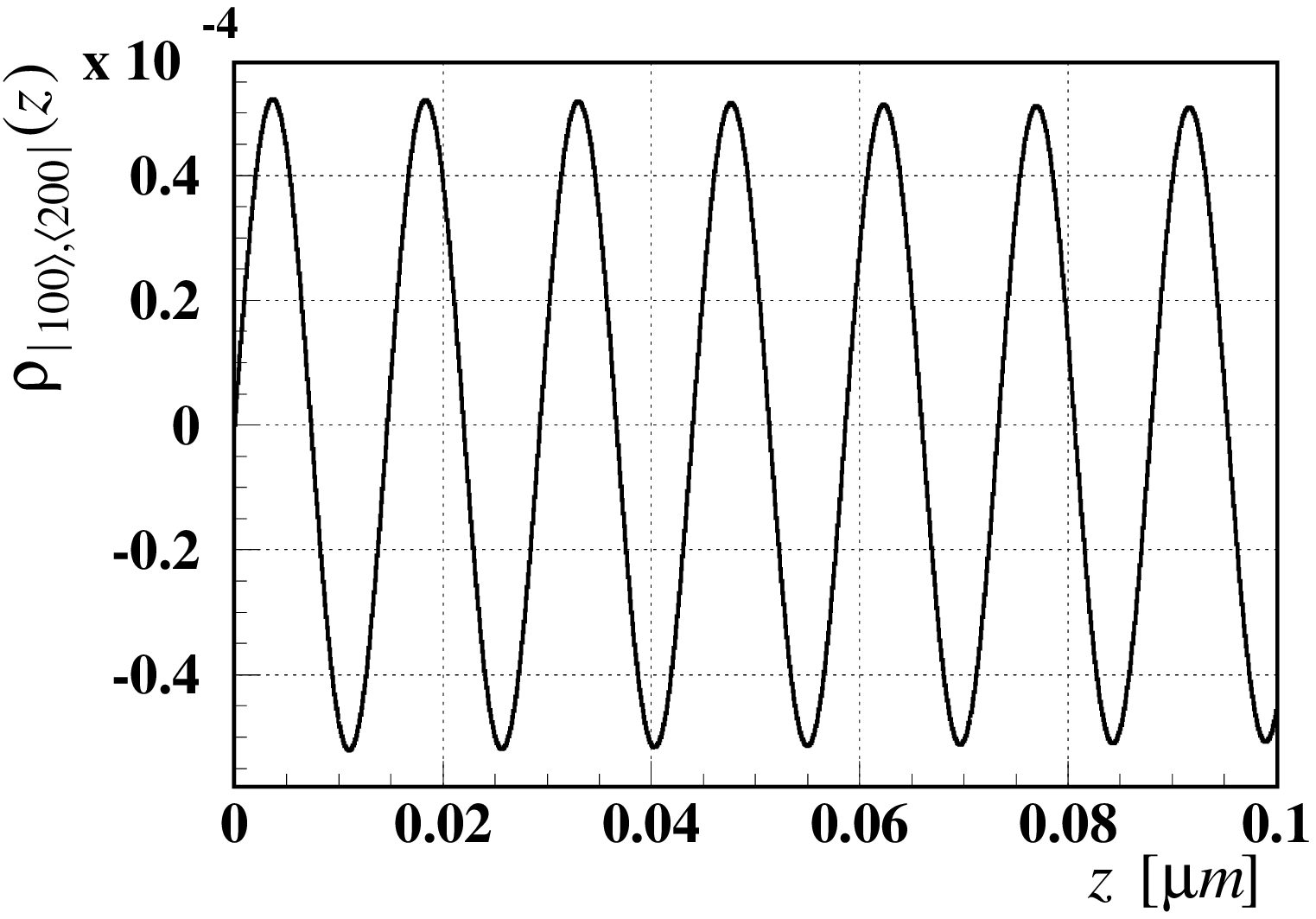}
\includegraphics[width=0.45\textwidth]{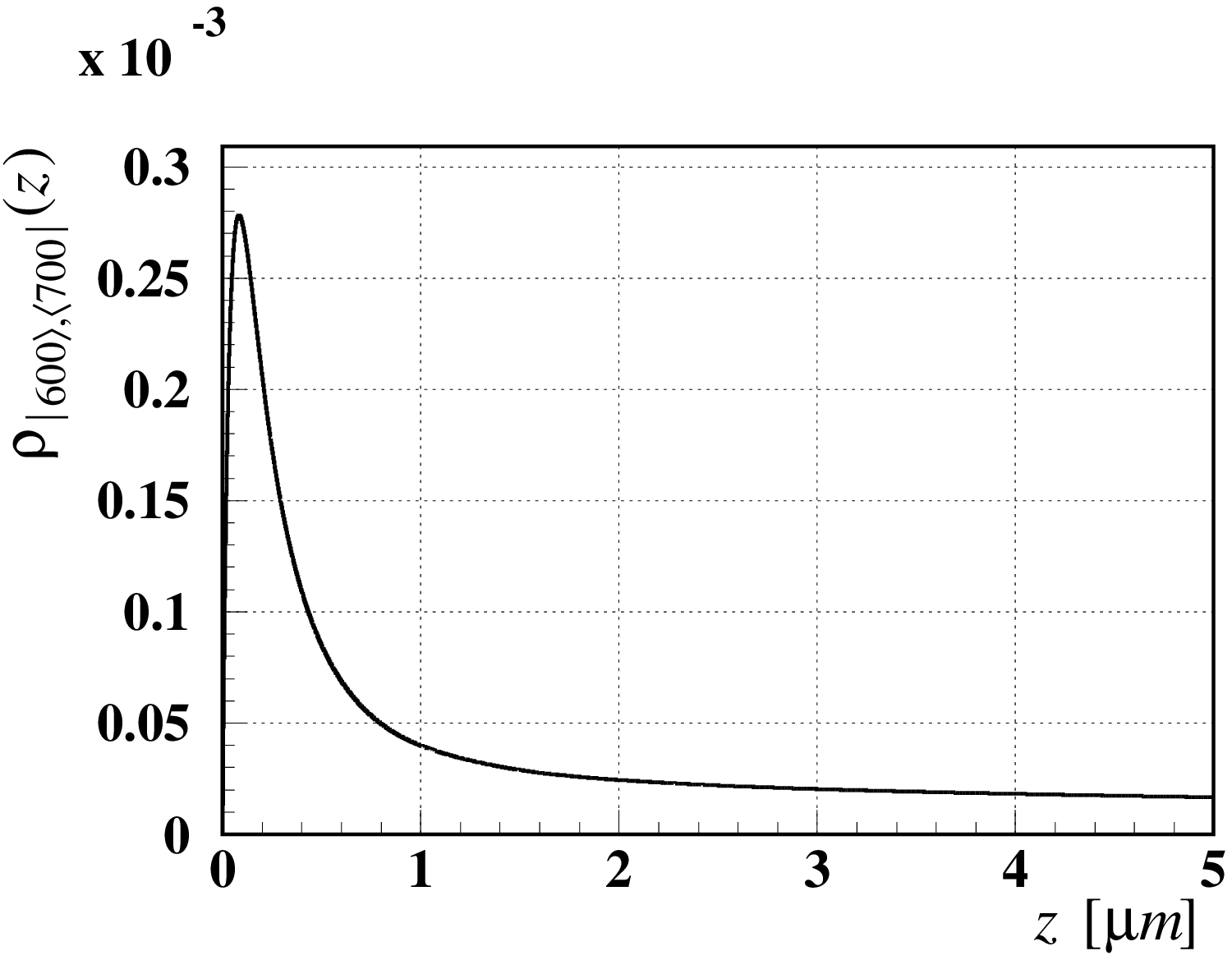}}
\caption{\label{fig:rhoij}The plots shows the solution of~\eref{eq:dedm} for
$\rho_{|100\rangle \langle 200|}$, oscillating with high frequency around 0,
and $\rho_{|600\rangle \langle 700|}$, constantly over 0 in a much larger range.}
\end{figure}

In~\fref{fig:rhoij} we can see that whereas $\rho_{ik}(z)$ oscillates
more than six times in $0.1$ $\mu m$ for $\rho_{|100\rangle\langle 200|}$ it does not oscillate
at all for $\rho_{|600\rangle\langle 700|}$ in a wide range.


As an example of the matrix elements we consider the subspace formed by the $|211\rangle$,
$|300\rangle$ and $|320\rangle$ states. The $\Omega$ matrix
restricted to this subspace is shown in~\tref{tab:omega}.
We can see that at least for the $|320\rangle \langle 300|$ mixed
state the matrix elements are of the same order of magnitude as for the
same shell pure states.

\begin{table}
\begin{center}
\caption{\label{tab:omega} $\Omega$ matrix elements in the $|211\rangle$, $|300\rangle$,
$|320\rangle$ subspace. Units are $10^{-20}$~barn.}
\begin{tabular}{c| c c c c c c c c} 
$\Omega_{|i\rangle \langle k|,|l\rangle \langle m|}$
  &$|211\rangle\langle211|$&$|300\rangle \langle 300|$&$|320\rangle \langle 300|$&$|320\rangle \langle 320|$\\
\hline
$|211\rangle \langle 211|$ & $-4.66$  & $0.044$ & $-0.083$ & $0.234$ \\
$|300\rangle \langle 300|$ & $0.044$  &$-18.2$  & $2.41$ & $0.$                 \\
$|320\rangle \langle 300|$ & $-0.083$& $2.41$  & $-13.7$ & $2.41$ \\
$|320\rangle \langle 320|$ & $0.234$  & $0.$    & $2.41$ &$-9.10$ 
\end{tabular} 
\end{center}
\end{table}

\section{Solving the System}

We have numerically solved the differential equation systems~\eref{eq:cl} and~\eref{eq:dedm} using
the Runge-Kutta method~\cite{nr}. Finding
the eigenvalues, as in~\cite{afan}, would be too lengthy due to the size of the density matrix system. The
Monte-Carlo method of~\cite{sant2} cannot be applied to the density matrix formalism
since the system~\eref{eq:dedm} {\bf does not obey}:
\begin{equation}
\Omega_{ik,ik} \ge \sum_{lm\neq ik} |\Omega_{ik,lm}|.
\end{equation}

We have considered a Nickel 95 $\mu m$ target and a monochromatic 4.6 $GeV/c$ atom sample. The
lifetime of the ground state of pionium was supposed to be $\Gamma^{-1}=2.9\cdot 10^{-15}$~s
according to the theoretical prediction~\cite{cola}.
The initial conditions are given by:
\begin{eqnarray}
P_i(0) = \rho_{ii}(0) = n_i^{-3}/\zeta(3)\qquad & \mbox{if $l_i=0$,}\nonumber \\
\label{eq:ic}
P_i(0) = \rho_{ij}(0) = 0\qquad & \mbox{otherwise,}
\end{eqnarray}
here $\zeta(3)=\sum n^{-3} \approx 1.202$. The system has been restricted to the bound
states with $n \le 7$. This means 84 mixed states and 353088 $\Omega$ matrix elements
different from zero. Cutting the number of considered states does only slightly affect the
solution of the last two cores taken into account (in this case states with $n=6$ and $n=7$) as
shown in~\cite{sant2}.

To achieve a very good accuracy in the final results we have considered a sequence of
step lengths in the numerical integration of the system:
\[ h= 2\cdot 10^{-3}, 1\cdot 10^{-3}, 0.5 \cdot 10^{-3}, 0.25\cdot 10^{-3}, 0.1\cdot 10^{-3} [\mu m] \]
and made a polynomial extrapolation to the limit $h=0$~\cite{nr}.

As we will explain below we are mainly interested in the averaged integrals of $\rho_{ii}(z)$ and $P_i(z)$ over the
target thickness $W$:
\begin{equation}
\label{eq:pdsci}
P_{dsc}^{i}= \frac{\int_{0}^{W} \rho_{i,i}(z) \rmd z}{W}.
\end{equation}
The $P$ picture in this equation is restored by changing $\rho_{ii}(z)\rightarrow P_i (z)$.
In~\tref{tab:pdsc} the $P_{dsc}$ results are shown as a function of the principal and angular
quantum number summed over the magnetic quantum number $m$ for a 95 $\mu m$ Nickel target.
The differences are not very large, especially for the ground and lowest excited states. However,
for some particular states the difference can be up to $20\%$. In~\fref{fig:probs} we see the
discrepancy for the case of the $|320\rangle$ state. 

\begin{table}
\begin{center}
\caption{\label{tab:pdsc} Summed $P_{dsc}^{nl}=\sum_m  P_{dsc}^{nlm}$ results in the probabilistic ($P$)
and density matrix ($\rho$) pictures. The average is over $W=95$ $\mu m$ and the target material is Nickel.}
{\footnotesize
\begin{tabular}{|c|c|l l l l l l|}
\hline 
 $P_{dsc}^{nl}$ & $P/\rho$ &   l=0     &    l=1      &     l=2     &    l=3      &     l=4     &    l=5      \\
\hline 
 n=1 & $P$    & 0.072854    &             &             &             &             &             \\
     & $\rho$ & 0.072860    &             &             &             &             &             \\
\hline 
 n=2 & $P$    &  0.0050676  & 0.008500    &             &             &             &             \\
     & $\rho$ &  0.0050878  & 0.008538    &             &             &             &             \\
\hline 
 n=3 & $P$    & 0.00087163  & 0.0016366   & 0.0020617   &             &             &             \\
     & $\rho$ & 0.00086909  & 0.0017234   & 0.0020250   &             &             &             \\
\hline 
 n=4 & $P$    &  0.00024899  & 0.0004803  & 0.0006270   & 0.0007326   &             &             \\
     & $\rho$ &  0.00024620  & 0.0005242  & 0.0006445   & 0.0007028   &             &             \\
\hline 
 n=5 & $P$    & 0.000092377 & 0.00018015  & 0.00023838  & 0.00028247  & 0.00031200  &             \\ 
     & $\rho$ & 0.000089072 & 0.00019899  & 0.00025137  & 0.00027925  & 0.00029343  &             \\ 
\hline 
 n=6 & $P$    & 0.000038357 & 0.000075133 & 0.000099834 & 0.00011906  & 0.000131889 & 0.00014113  \\
     & $\rho$ & 0.000034640 & 0.000079850 & 0.000102429 & 0.00011493  & 0.000121041 & 0.00012520  \\
\hline 
 n=7 & $P$    & 0.000015300 & 0.000029939 & 0.000039634 & 0.000047316 & 0.000052490 & 0.000057376 \\ 
     & $\rho$ & 0.000013706 & 0.000031028 & 0.000039316 & 0.000043479 & 0.000048089 & 0.000045044 \\ 
\hline
\end{tabular}
}
\end{center}
\end{table}

\begin{figure}
\begin{center}
\includegraphics[width=8cm]{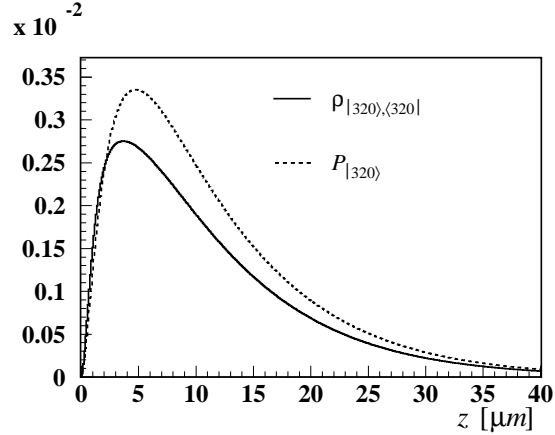}
\end{center}
\caption{\label{fig:rho320} The solution of~\eref{eq:cl} and~\eref{eq:dedm} for the $|320\rangle$ state.}
\end{figure}

\subsection{Obtaining the Break-up Probability}

Our goal is to obtain the break-up probability ($P_{br}$) of pionium in
the target. As we have seen in the previous sections the atoms in the target can suffer transitions between bound
states and annihilate. However, they can also be transferred, in a collision with a target atom, into a
continuum state. The coefficients $c_{i,l}$ and $\Omega_{ik,lm}$ accounting for transitions between discrete and continuum states
are more difficult to compute than the discrete-discrete ones since the atomic form factors have a more
complicate expression~\cite{hala}.
However, as shown in~\cite{sant2} for the case of the probabilistic picture, the direct calculation of break-up probability
from the systems solutions is not satisfactory since it decreases very slowly as a function of the principal quantum
number of the broken discrete state and only a finite number of shells ($n\le 7$) are considered
when solving either~\eref{eq:cl} and~\eref{eq:dedm}. We would have to guess the break up probability for any shell with
$n>7$ and make a large error in the total break-up probability determination.

The standard strategy
to obtain break-up probability consists of calculating the probability of the atom to leave the target in
a discrete state ($P_{dsc}$) and the probability of annihilation ($P_{anh}$) and make use of the relation:
\begin{equation}
1=P_{br} + P_{dsc} + P_{anh}.
\end{equation}
As both $P_{dsc}$ and $P_{anh}$ quickly decrease with $n$ we have an accurate result taking into account
only those events with $n \le 7$. A small correction will be introduced for $P_{dsc}^{n>7}$.

In the experimental conditions the atoms are not created at the target beginning but uniformly distributed
along the target thickness. The probability that the atom leaves the target in a discrete state can be however linked
to the solutions under~\eref{eq:ic} initial conditions by:
\begin{equation}
\label{eq:pdscs}
P_{dsc} = \sum_{i} \frac{\int_{0}^{W} \rho_{ii}(W-z) \rmd z}{W}=\sum_{i} \frac{\int_{0}^{W} \rho_{ii}(z) \rmd z}{W}
\end{equation}
where $W$ stands for the target thickness (of 95 $\mu m$ in our case). 

The annihilation probability is a little bit more difficult to calculate. If the atom
is created in $z_0$, the probability that it flies to $z$ and annihilates is given by $\Gamma_i\rho_{ii}$.
But $z$ can be any value between $z_0$ and the target end $W$. Meanwhile, the atom is randomly created
between 0 and $W$ with uniform distribution, then the annihilation probability is given by:
\begin{equation}
\label{eq:panns}
P_{anh} = \sum_i \frac{\Gamma_i}{W} \int_0^{W} \int_{z_0}^{W} \rho_{ii}(z-z_0) \rmd z \rmd z_0 =
\sum_i \frac{\Gamma_i}{W} \int_0^{W} (W-z) \rho_{ii}(z) \rmd z
\end{equation}
Of course the probabilistic picture is
restored by substituting $\rho_{ii}(z)$ by $P_i(z)$ in~\eref{eq:pdscs} and~\eref{eq:panns}.

As we did in~\eref{eq:pdsci} for the $P_{dsc}^i$ probability we can of course define the annihilation probability
from a certain state as:
\begin{equation}
P_{anh}^i= \frac{\Gamma_i}{W} \int_0^{W} (W-z) \rho_{ii}(z) \rmd z
\end{equation}
where again the replacement $\rho_{ii}(z)\rightarrow P_i(z)$ recovers the $P$ picture. Of course
$P_{anh}^i=0$ for any state with $l_i \neq 0$.

The results for the annihilation probability from the $S$ states up to $n=7$ are shown in~\tref{tab:panh}
and complete those of the $P_{dsc}$ in~\tref{tab:pdsc}.

\begin{table}
\begin{center}
\caption{\label{tab:panh} $P_{anh}^{n}$ results in the $P$ and $\rho$ pictures.
The average is over $W=95$ $\mu m$ and the target material is Nickel. The lifetime of pionium
was assumed to be $2.9 \times 10^{-15}$~s.}
\begin{tabular}{|c|c|c|}
\hline
 $n$ & $P_{anh}^{n}$ & $P/\rho$  \\
\hline
 $n=1$  & $P$    & 0.072854    \\
     & $\rho$ & 0.072860    \\
\hline
 $n=2$ & $P$    &  0.0050676  \\
     & $\rho$ &  0.0050878  \\
\hline
 $n=3$ & $P$    & 0.00087163  \\
     & $\rho$ & 0.00086909  \\
\hline
 $n=4$ & $P$    &  0.00024899 \\
     & $\rho$ &  0.00024620 \\
\hline
 $n=5$ & $P$    & 0.000092377 \\
     & $\rho$ & 0.000089072 \\
\hline
 $n=6$ & $P$    & 0.000038357 \\
     & $\rho$ & 0.000034640 \\
\hline
 $n=7$ & $P$    & 0.000015300 \\
     & $\rho$ & 0.000013706 \\
\hline
\end{tabular}
\end{center}
\end{table}

\begin{figure}
\begin{center}
\includegraphics[width=8cm]{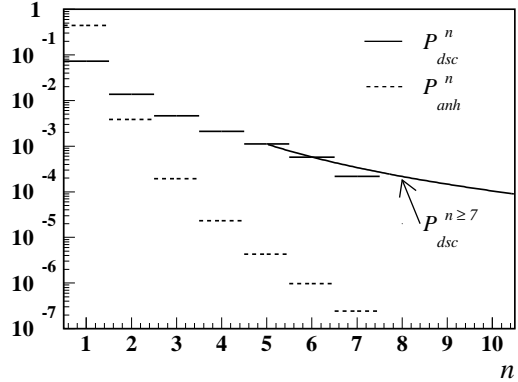}
\end{center}
\caption{\label{fig:probs} Dependence of $P^{i}_{dsc}$ and $P^{i}_{anh}$ averaged
over every shell on the principal quantum number. The extrapolation of~\eref{eq:extr}
is also shown.}
\end{figure}

In~\fref{fig:probs} we can see the dependence of $P^{n}_{dsc}$ and $P^{n}_{anh}$ on the principal quantum number. The
results have been summed over every shell bound states. We can check that whereas $P^{n}_{anh}$ quickly converges to
zero, and can be neglected for $n_i > 4$, $P^{i}_{dsc}$ diminishes more slowly. This leads to introduce an
extrapolation for $P^{n\geq7}_{dsc}$~\cite{afan}:
\begin{equation}
\label{eq:extr}
P^{n\geq7}_{dsc} = \frac{a}{n^3} + \frac{b}{n^5},
\end{equation}
where $a$ and $b$ are obtained by fitting $P^{n}_{dsc}$ at $n=5$ and $n=6$. The extrapolation is also used
for $n=7$ because not considering the next shells in the systems distorts this shell solutions.

The extrapolation results are summed over $n$ and, together with $P^{n<7}_{dsc}$ and $P_{anh}$, subtracted to
one to calculate the break-up probability:
\begin{equation}
P_{br} = 1 - P_{anh} - P^{n_i<7}_{dsc} - P^{n_i \geq 7}_{dsc}
\end{equation}
obtaining, for out particular example of $2.9 \times 10^{-15}$ $s$ atoms in a Ni 25 $\mu m$ target: $P_{br}=0.459254$
in the probabilistic picture and $P_{br}=0.459268$ in the density matrix formalism. The other probabilities are shown
in~\tref{tab:probs}.

\begin{table}
\begin{center}
\caption{\label{tab:probs} Probability results in the $P$ and $\rho$ pictures.
The average is over $W=95$ $\mu m$ and the target material is Nickel. The lifetime of pionium
was assumed to be $2.9 \times 10^{-15}$ $s$.}
\begin{tabular}{ccccc}
\hline
Picture & $P_{br}$ & $P_{anh}$ & $P^{n<7}_{dsc}$ & $P^{n \geq 7}_{dsc}$ \\
\hline
  $P$   & 0.459254 & 0.444536 & 0.0947916 & 0.001418 \\
$\rho$  & 0.459268 & 0.444575 & 0.0949106 & 0.001245 \\
\hline
\end{tabular}
\end{center}
\end{table}

\section{Discussion and Conclusions}

We have checked that in the conditions of the DIRAC experiment the effect of the quantum
interference between states does not change the result of the break-up probability
of pionium in the target. Hence, the results obtained in the classical picture are accurate
enough to safely perform the experimental measurement.

The unchanged result of break-up result takes place despite the fact that for some
discrete states, as $|320\rangle$, the effect of interference can significantly change
the population of the state up to $20\%$ levels. However, the most affected states
are very unpopulated and hence not relevant for the final results.

The situation could change if the initial conditions were not that most atoms are created
in the ground state. The later is non degenerated and interferences only show-up
after a first transition. However we have checked
what would happen if the initial conditions were that all the atoms were created in the
$|300\rangle$ state and neither found a significant change with the probabilistic
approach. A possible explanation is that while the interference is most likely
with states with the same magnetic quantum number $m$, and
comparable with the transition cross sections, the dominant transitions are those
that increase $l$ and $m$ in one unit, free of interference with the father state. 

\ack

We would like to thank L Nemenov and L Tauscher D Trautmann for their support. We would also like
to thank K Hencken for his help. L. Afanasyev, A Tarasov and O Voskresenskaya would like
to acknowledge the interesting discussions during the workshop HadAtom03, partially supported
by the ECT.

\appendix

\section{The $\Omega$ matrix elements in the First Born Approximation}
\label{app:omega}

Let us show how to obtain the discrete matrix elements of the $\Omega$
matrix in the first Born approximation from the original equations
of~\cite{voskre}. The $\Omega$ operator is originally defined as a function of the transverse
position of the atom wave functions $\vec{s}_{1,2}$.
If we split the operator in two:
\begin{equation}
  \label{eq:1}
\Omega(\vec{s_1},\vec{s_2}) = \Omega^{(1)}(\vec{s_1},\vec{s_2}) +
                             \Omega^{(2)}(\vec{s_1},\vec{s_2}),
\end{equation}
its definition will be given by:
\begin{eqnarray}
  \label{eq:2}
  \Omega^{(1)}(\vec{s_1},\vec{s_2}) & = &\int
  \left[\Gamma(\vec{b},\vec{s_1}) + \Gamma^*(\vec{b},\vec{s_2})\right] \rmd^2b, \\
  \label{eq:3}
  \Omega^{(2)}(\vec{s_1},\vec{s_2}) & = & - \int
  \Gamma(\vec{b},\vec{s_1})\Gamma^*(\vec{b},\vec{s_2}) \: \rmd^2b.
\end{eqnarray}
In the case of the $\pi^+\pi^-$-atom the interaction operator
of the Glauber theory is given by:
\begin{equation}
  \label{eq:4}
\Gamma(\vec{b},\vec{s}) = 1 - \exp\left[ i\chi(\vec{b}-\vec{s}/2) -
  i\chi(\vec{b}+\vec{s}/2)\right],
\end{equation}
where
\begin{equation}
  \label{eq:5}
\chi(\vec{B}) = \frac{1}{\beta}\int\limits_{-\infty}^{\infty} U(\sqrt{B^2+z^2}) \:\rmd z,
\end{equation}
being $U(r)$ the potential of the target atoms given by the inverse Fourier transform
of~\eref{eq:pot}.

First of all we are going to re-write $\Omega^{(1)}(\vec{s_1},\vec{s_2})$. For that we split $\Gamma(\vec{b},\vec{s})$
into its real and imaginary part:
\begin{equation}
  \label{eq:7}
\Gamma(\vec{b},\vec{s}_{1(2)}) =
\mathrm{Re}\:\Gamma(\vec{b},\vec{s}_{1(2)}) +
i\:\mathrm{Im}\:\Gamma(\vec{b},\vec{s}_{1(2)}),
\end{equation}
\begin{eqnarray}
  \label{eq:8}
\mathrm{Re}\:\Gamma(\vec{b},\vec{s}_{1(2)}) & = & 1 -\cos\left[
\chi(\vec{b}-\vec{s}/2)-\chi(\vec{b}+\vec{s}/2)\right] \nonumber \\
&=& \frac{1}{2} \Gamma(\vec{b},\vec{s}_{1(2)}) \: \Gamma^*(\vec{b},\vec{s}_{1(2)}),
\end{eqnarray}
\begin{equation}
  \label{eq:8a}
\mathrm{Im}\:\Gamma(\vec{b},\vec{s}_{1(2)}) = -\sin\left[
\chi(\vec{b}-\vec{s}/2)-\chi(\vec{b}+\vec{s}/2)\right],
\end{equation}
where the integral over the imaginary part goes to zero:
\begin{equation}
  \label{eq:9}
\int \mathrm{Im}\:\Gamma(\vec{b},\vec{s}_{1(2)}) \: \rmd^2 b = 0,
\end{equation}
due to the odd nature of the $\sin$ function and the even nature of $\chi(\vec{b}\pm \vec{s}/2)$.
Taking this into account we can have:
\begin{equation}
  \label{eq:10}
\Omega^{(1)}(\vec{s_1},\vec{s_2})= \frac{1}{2}
\int 
\left[\Gamma(\vec{b},\vec{s}_{1})\Gamma^*(\vec{b},\vec{s}_{1})+
\Gamma(\vec{b},\vec{s}_{2})\Gamma^*(\vec{b},\vec{s}_{2})\right]\: \rmd^2 b.
\end{equation}

Our final goal is to obtain the matrix elements $\Omega^{(1,2)}_{ik,\;lm}$ defined as:
\begin{equation}
  \label{eq:11}
\Omega^{(1,2)}_{ik,\;lm}=\int \psi_i^*(\vec{r}_{1})
\psi_l(\vec{r}_{1})
\psi_k(\vec{r}_{2})
\psi_m^*(\vec{r}_{2})\Omega^{(1,2)}(\vec{s_1},\vec{s_2}) \:\rmd\vec{r}_{1}\:\rmd\vec{r}_{2}.
\end{equation}
In particular we can define the profile-function $\Gamma_{il}(\vec{b})$:
\begin{equation}
  \label{eq:12}
\Gamma_{il}(\vec{b}) = \int 
\psi_i^*(\vec{r}) \psi_l(\vec{r}) \Gamma_{il}(\vec{b},\vec{s}) \: \rmd\vec{r},
\end{equation}
and its Fourier transform, the amplitude:
\begin{equation}
  \label{eq:13}
A_{il}(\vec{q}) = \frac{i}{2\pi} \int e^{i\vec{q}\vec{b}} \Gamma_{il}(\vec{b}) \: \rmd^2 b,
\end{equation}
\begin{equation}
  \label{eq:14}
\Gamma_{il}(\vec{b}) = \frac{1}{2\pi i} \int e^{-i\vec{q}\vec{b}} A_{il}(\vec{q}) \: \rmd^2 q.
\end{equation}
It is easy to check that:
\begin{equation}
  \label{eq:omega2}
\Omega^{(2)}_{ik,\;lm}=- \int \Gamma_{il}(\vec{b})\:\Gamma_{km}^*(\vec{b}) \: \rmd^2 b = - \int 
A_{il}(\vec{q})\:A_{km}^*(\vec{q}) \: \rmd^2 q.
\end{equation}
To obtain an analogue of~\eref{eq:omega2} for $\Omega^{(1)}_{ik,\;lm}$ we have to work a little bit.
Of course, by definition:
\begin{eqnarray}
  \label{eq:omega1}
\Omega^{(1)}_{ik,\;lm} & = &\frac{\delta_{km}}{2} \int \left[\int \psi_i^*(\vec{r}) \psi_l(\vec{r})
 \Gamma(\vec{b},\vec{s})\:\Gamma^*(\vec{b},\vec{s}) \:\rmd\vec{r}\right] \: \rmd^2 b \nonumber \\
&&{}+\frac{\delta_{il}}{2} \int \left[\int 
 \psi_k^*(\vec{r}) \psi_m(\vec{r})
 \Gamma(\vec{b},\vec{s})\:\Gamma^*(\vec{b},\vec{s})\:\rmd\vec{r} \right]^* \: \rmd^2 b.
\end{eqnarray}
To achieve the final result we will need the completeness equation in the form:
\begin{equation}
  \label{eq:delta}
\delta(\vec{r}-\vec{r'})=\sum_j \psi_j(\vec{r}) \psi_j^*(\vec{r'}),
\end{equation}
which allows to express the inner integrals in~\eref{eq:omega1} in terms of the
profile-function $\Gamma_{ij}(\vec{b})$:
\begin{eqnarray}
\fl \label{eq:int1} \int \psi_i^*(\vec{r}) \psi_l(\vec{r})
 \Gamma(\vec{b},\vec{s})\:\Gamma^*(\vec{b},\vec{s}) \: \rmd\vec{r} = 
 \int \psi_i^*(\vec{r}) \psi_l(\vec{r'})
\delta(\vec{r}-\vec{r'}) \Gamma(\vec{b},\vec{s})\:\Gamma^*(\vec{b},\vec{s'}) \: \rmd\vec{r} \: \rmd\vec{r'}\nonumber \\
\fl = \sum_j \left[\int \psi_i^*(\vec{r})\Gamma(\vec{b},\vec{s})
 \psi_j(\vec{r})\:\rmd\vec{r}\right] \left[\int \psi_j^*(\vec{r'})\Gamma^*(\vec{b},\vec{s'})
 \psi_l(\vec{r'}) \: \rmd\vec{r'} \right]  = \sum_j \Gamma_{ij}(\vec{b})\:\Gamma_{lj}^*(\vec{b}),
\end{eqnarray}
where of course we can make the substitution:
\begin{equation}
  \label{eq:int2}
\int \sum_j
  \Gamma_{ij}(\vec{b})\:\Gamma_{lj}^*(\vec{b})\:\rmd^2 b =
\int \sum_j A_{ij}(\vec{q}) A_{lj}^*(\vec{q}) \:\rmd^2 q,
\end{equation}
to obtain:
\begin{equation}
\Omega^{(1)}_{ik,\;lm} = \frac{\delta_{km}}{2} \int \sum_j A_{ij}(\vec{q}) A_{lj}^*(\vec{q}) \:\rmd^2 q
+\frac{\delta_{il}}{2} \int \sum_j A_{mj}(\vec{q}) A_{kj}^*(\vec{q}) \:\rmd^2 q.
\end{equation}

In the Born approximation
\begin{equation}
  \label{eq:a_born}
A_{il}(\vec{q}) = \frac{1}{\beta} U(\vec{q})
\left[ F_{i}^{l}\left(\frac{\vec{q}}{2}\right)- F_{i}^{l}\left(-\frac{\vec{q}}{2}\right)\right],
\end{equation}
where we find the form factors defined in~\eref{eq:ff}.

Let us try to perform the sum:
\begin{eqnarray}
  \label{eq:sum_s}
\fl \sum_j A_{ij}(\vec{q}) A_{lj}^*(\vec{q})= \sum_j
\left[ F_{i}^{j}\left(\frac{\vec{q}}{2}\right) - F_{i}^{j}\left(-\frac{\vec{q}}{2}\right)\right]
\left[ F_{l}^{j}\left(\frac{\vec{q}}{2}\right) - F_{l}^{j}\left(-\frac{\vec{q}}{2}\right)\right]^* \nonumber\\
\fl = \sum_j \left[ \int \psi_i^*(\vec{r})
\left( e^{i\vec{q}\vec{r}/2} - e^{- i\vec{q}\vec{r}/2}\right)
\psi_j(\vec{r}) \: \rmd\vec{r}\right]
\left[ \int \psi_j^*(\vec{r'})
\left( e^{-i\vec{q}\vec{r'}/2} - e^{i\vec{q}\vec{r'}/2}\right)
\psi_l(\vec{r'}) \:\rmd\vec{r'} \right]
\nonumber\\
\fl = \int \psi_i^*(\vec{r})
\left( e^{i\vec{q}\vec{r}/2} - e^{- i\vec{q}\vec{r}/2}\right)
\left( e^{-i\vec{q}\vec{r}/2} - e^{i\vec{q}\vec{r}/2}\right)
\psi_l(\vec{r}) \: \rmd\vec{r}
\nonumber\\
\fl = \int \psi_i^*(\vec{r})
\left( 2 - e^{-i\vec{q}\vec{r}} - e^{i\vec{q}\vec{r}/2}\right)
\psi_l(\vec{r}) \: \rmd\vec{r}
= 2\delta_{il} - F_{i}^{l}(\vec{q}) - F_{i}^{l}(-\vec{q}).
\end{eqnarray}

From equations~(\ref{eq:omega2}), (\ref{eq:omega1}), (\ref{eq:int1}),
(\ref{eq:int2}), (\ref{eq:a_born}) and (\ref{eq:sum_s}) one
can derive the final expressions in the Born approximation:
\begin{equation}
  \label{eq:omega_b}
\Omega_{ik,\;lm} = \Omega^{(1)}_{ik,\;lm} + \Omega^{(2)}_{ik,\;lm},
\end{equation}
\begin{eqnarray}
\Omega^{(1)}_{ik,lm} =
& \frac{\delta_{k,m}}{2\beta^2}  \int
\left| U(q) \right|^{2} \left[2 \delta_{i,l}-F_{i}^{l}(\vec{q})-F_{i}^{l}(-\vec{q})
\right] \rmd^2 q +  \nonumber \\
& + \frac{\delta_{i,l}}{2\beta^2} \int
\left| U(q) \right|^{2} \left[2 \delta_{k,m}-F_{k}^{m}(\vec{q})-F_{k}^{m}(-\vec{q})
\right] \rmd^2 q, \;\;\;\;\;\;\;\;\;\;\;\;\;\;\;\;\;\;\; \mbox{\eref{eq:om1}} \nonumber
\end{eqnarray}
\begin{eqnarray}
\Omega^{(2)}_{ik,lm} = \frac{1}{\beta^2}
  \int
 \left| U(q) \right|^{2} & \left[
F_{i}^{l}\left(\frac{\vec{q}}{2}\right)-
F_{i}^{l}\left(-\frac{\vec{q}}{2}\right) \right] \times \nonumber \\
&\times \left[F_{k}^{m}\left(\frac{\vec{q}}{2}\right)-
F_{k}^{m}\left(-\frac{\vec{q}}{2}\right) \right]^{*} \rmd^2 q.
\;\;\;\;\;\;\;\;\;\;\;\;\;\;\;\;\;\;\;\;\;\;\;\;\; \mbox{\eref{eq:om2}} \nonumber
\end{eqnarray}

\end{document}